\documentclass[reprint,twocolumn]{revtex4}
\usepackage{amsfonts}
\usepackage{amsmath}
\usepackage{amssymb}
\usepackage{charter}
\usepackage{dcolumn}
\usepackage{graphicx}
\usepackage{float}

\begin{document}

\title{Evaluating the quantum optimal biased bound in a unitary evolution
process}
\author{Shoukang Chang$^{1}$}
\author{Wei Ye$^{2}$}
\thanks{Corresponding author. 71147@nchu.edu.cn}
\author{Xuan Rao$^{2}$}
\author{Huan Zhang$^{3}$}
\author{Liqing Huang$^{1}$}
\author{Mengmeng Luo$^{4}$}
\author{Yuetao Chen$^{1}$}
\author{Qiang Ma$^{1}$}
\author{Shaoyan Gao$^{1}$}
\thanks{Corresponding author. gaosy@xjtu.edu.cn}
\affiliation{$^{{\small 1}}$\textit{MOE Key Laboratory for Nonequilibrium Synthesis and
Modulation of Condensed Matter, Shaanxi Province Key Laboratory of Quantum
Information and Quantum Optoelectronic Devices, School of Physics, Xi'an
Jiaotong University, Xi'an 710049, People's Republic of China}\\
$^{{\small 2}}$\textit{School of Information Engineering, Nanchang Hangkong
University, Nanchang 330063, China}\\
$^{{\small 3}}${\small \ }\textit{School of Physics, Sun Yat-sen University,
Guangzhou 510275, China}\\
$^{4}$\textit{Department of Physics, Xi'an Jiaotong University City College,
Xi'an 710018, China}}

\begin{abstract}
Seeking the available precision limit of unknown parameters is a significant
task in quantum parameter estimation. One often resorts to the widely
utilized quantum Cram\'{e}r-Rao bound (QCRB) based on unbiased estimators to
finish this task. Nevertheless, most actual estimators are usually biased in
the limited number of trials. For this reason, we introduce two effective
error bounds for biased estimators based on a unitary evolution process in
the framework of the quantum optimal biased bound. Furthermore, we show
their estimation performance by two specific examples of the unitary
evolution process, including the phase encoding and the SU(2) interferometer
process. Our findings will provide an useful guidance for finding the
precision limit of unknown parameters.

\end{abstract}

\maketitle

\section{Introduction}

Quantum parameter estimation is a burgeoning interdisciplinary field between
classical parameter estimation theory and quantum mechanics, \cite{1,2,3}
and has attracted a great deal of attention in recent decades since it not
only plays a core role in \textquotedblleft the second quantum
revolution\textquotedblright \  \cite{4} but also has many prominent
applications, such as the search of dark matters \cite{5}, optical imaging
\cite{6,7}, and gravitational wave detection \cite{8}. The primary goal of
such an estimation theory is to actualize high-precision measurement of
unknown parameters than its classical counterparts by designing estimation
schemes with some quantum effects, such as entanglement \cite{9,10,11},
nonclassicality \cite{12}, and quantum correlation \cite{13,14,15}. To
differentiate between excellent and worse estimation schemes, one usually
invokes some error bounds to describe the theoretical precision limit of the
estimation scheme \cite{16,17}. The most famous mathematical bound is the
quantum Cram\'{e}r-Rao bound (QCRB) proposed by Helstrom \cite{18}, which
has been widely studied and used in many research fields, such as quantum
sensing \cite{19,20,21}, quantum illumination \cite{22,23}, and
quantification of nonclassicality \cite{24,25,26}. Especially, M. Tsang
\emph{et al}. investigated the ultimate precision limit of locating two weak
thermal optical point sources in the framework of the QCRB \cite{6}. It is
worth noting that the QCRB based on the unbiased estimator is reasonable and
provides the asymptotically tight precision limit when the number of
repeated trials is sufficiently huge \cite{27,28,29,30}. Hence, the QCRB can
hardly offer the attainable accuracy limit of unknown parameters with a
limited number of trials \cite{29,30}.

For this purpose, one has proposed many new error bounds via the mini-max
approach \cite{31,32}, binary hypothesis testing \cite{33,34,35}, and
optimal measurement strategy or estimator \cite{30,36,37,38}, which are
alternatives to the QCRB for single-(or multi-) parameter estimation. For
the single-parameter estimation, M. Hayashi introduced the local asymptotic
mini-max bound which is strictly larger than QCRB for the phase estimation
\cite{31}. After that, M. Tsang put forward the quantum Ziv-Zakai bound and
further showed that the error bound can be superior to the QCRB for probe
states with highly non-Gaussian photon-number statistics \cite{33}. It is
noteworthy that such a bound can not be completely guaranteed to be superior
to the QCRB \cite{33,39}. In order to tackle this problem, X. M. Lu and M.
Tsang gave the quantum versions of the Weiss-Weinstein bounds and
demonstrated that the new error bound can be much tighter than both QCRB and
quantum Ziv-Zakai bound \cite{39}. Subsequently, Mart\'{\i}nez-Vargas \emph{%
et al}. presented a modification of the quantum Van Trees inequality, which
derived a new error bound by optimizing measurement strategy for estimating
the value of unknown parameters and indicated that the error bound can
provide better estimation performance than conventional QCRB \cite{36}.
Likewise, for the multi-parameter estimation, there also exist some
non-asymptotic error bounds as alternatives to the commonly utilized QCRB
\cite{34,40,41,42,43,44,45,46}. One of these error bounds was given in Ref.
\cite{34}, which proposed the quantum Bell-Ziv-Zakai Bound for the
multi-parameter estimation and further investigated the quantum optical
phase waveform estimation problem. Additionally, J. Rubio and J. Dunningham
derived a quantum Bayesian multi-parameter bound with limited data and
demonstrated that the new multi-parameter bound is better than the QCRB for
a qubit sensing network \cite{41}.

On the other hand, the quantum optimal biased bound (QOBB) is one of error
bounds which can offer lower precision limit than QCRB under a finite number
of measurements \cite{47}. The QOBB is a efficacious lower bound for all
estimators (biased or unbiased), which can also be used as the criterion for
quantum parameter estimation \cite{47}. We also noticed that the QOBB is
obtained by the quantization of the classical optimal biased bound where the
quantization process invokes the symmetric logarithmic derivative, which
makes that the QOBB is related to the quantum Fisher information \cite{47}.
It is worth affirming that this quantization process is natural and
reasonable since the classical optimal biased bound is expressed in the
classical Fisher information \cite{48,49}. However, the quantization process
of the classical optimal biased bound is not unique. Therefore, in this
paper, we consider two other quantization processes to derive two new
versions of the QOBB, i.e., QOBB1 and QOBB2. These two novel QOBBs apply to
that the encoding process of quantum states for unknown parameters such as
the phase shift and field-quadrature displacement satisfies the unitary
evolution \cite{11,20,30,44}. We further show that the QOBB1 and QOBB2 can
provide lower precision limit than original QOBB in the case of initial
mixed state. For the initial pure state scenario, the value of QOBB1 is
still lower than QOBB while both QOBB2 and QOBB are numerically equal. It is
noteworthy that although the QOBB1 and QOBB2 are always not higher than QOBB
in numerical value, the tightness of these two new versions of the QOBB is
lower than that of the original QOBB. Moreover, we illustrate our results
with two specific examples of the unitary evolution process, including the
phase encoding and the SU(2) interferometer process.

This paper is arranged as follows. In Sec. II, we briefly summarize the
known results of both the minimum mean square error and the QOBB. In Sec.
III, we derive two new versions of the QOBB under the unitary evolution
process. In Sec. IV, we demonstrate our results with two specific examples
of the unitary evolution process, including the phase encoding and the SU(2)
interferometer process. Finally, the main conclusions are drawn in the last
section.

\section{Minimum mean square error and Quantum optimal biased bound}

To begin with, we briefly review the known results for parameter estimation
based on the minimum mean square error (MMSE) and QOBB. Let $x$\ denote the
unknown parameter to be estimated, and $\check{x}(y)$\ be an estimator of $x$%
\ from the observation results $y$. The estimation precision of the unknown
parameter $x$\ can be quantified through the mean square error, i.e.,%
\begin{equation}
\Sigma =\int p(x)p(y|x)(\check{x}(y)-x)^{2}dxdy,  \label{1}
\end{equation}%
where $p(x)$\ is the prior probability density and $p(y|x)$\ is the
condition observation probability density. According to Refs. \cite{30,38,47}%
, the classical lower bound for $\Sigma ,$\ i.e., MMSE, can be given by%
\begin{eqnarray}
MMSE &=&\int p(x)p(y|x)(\check{x}(y)_{opt}-x)^{2}dxdy  \notag \\
&=&\int p(x)x^{2}dx-\int p(y)\check{x}(y)_{opt}^{2}dy,  \label{2}
\end{eqnarray}%
where $\check{x}(y)_{opt}=\int xp(x|y)dx$\ is the optimal estimator of the
mean square error, $p(x|y)=p(x)p(y|x)\left/ p(y)\right. $\ is the posterior
probability density, and $p(y)=\int p(x)p(y|x)dx$\ is the marginal
probability density.

However, as the optimal estimator is often biased in limited trials, its
performance cannot be calibrated by the QCRB within a finite range \cite{47}%
. For this reason, J. Liu and H. D. Yuan proposed the QOBB, which is an
effective lower bound for all estimators \cite{47}. Next we will mainly
review this important lower bound for quantum parameter estimation.

When given the prior probability density $p(x)$ of the general unknown
function $f(x)$ with parameters $x$ to be estimated, and $\breve{f}(y)$ is
an estimator of the unknown function $f(x)$ from the observation results $y$%
. Accordingly, the expectation value of the estimator $\breve{f}(y)$ can be
expressed as
\begin{eqnarray}
E_{y|x}[\breve{f}(y)] &=&\int \breve{f}(y)p(y|x)dy  \notag \\
&=&f(x)+b_{0}(x),  \label{3}
\end{eqnarray}%
where $b_{0}(x)$ is the bias of the estimator $\breve{f}(y)$ and $p(y|x)$
represents the condition observation probability density. For the quantum
parameter estimation problem, the condition observation probability density $%
p(y|x)$\ turns out to be according to the Born's rule
\begin{equation}
p(y|x)=\text{Tr}(\hat{\rho}_{x}\hat{\Pi}_{y}),  \label{4}
\end{equation}%
where $\hat{\rho}_{x}$\ is the density operator encoding the unknown
function $f(x)$, $\hat{\Pi}_{y}$\ is the positive operator-valued measure,
and the symbol of Tr is the operator trace. On this background, the
mean-square estimation error is defined as%
\begin{equation}
\Sigma =\int [\breve{f}(y)-f(x)]^{2}p(x,y)dxdy,  \label{5}
\end{equation}%
where $p(x,y)=p(y|x)p(x)$ is the joint probability density. According to
Ref. \cite{47}, a lower bound for $\Sigma $ shown in Eq. (\ref{5}) can be
thus given by%
\begin{equation}
\Sigma \geq \int p(x)\left[ b_{0}^{2}(x)+(f^{\prime }(x)+b_{0}^{\prime
}(x))^{2}/F\right] dx,  \label{6}
\end{equation}%
where the superscript $\prime $ is the derivative with respect to $x$ and $F$%
\textbf{\ }is the quantum Fisher information. In order to achieve the valid
lower bound for all estimators, the variational method is often used to find
out the optimal $b_{0}(x)$ that minimizes the bound given in Eq. (\ref{6}).
To be more specific, when assuming the prior probability density $p(x)$\ is
limited at\textbf{\ }$0\leq x\leq a$ and denoting as $G(b_{0},x)=$\ $p(x)%
\left[ b_{0}^{2}(x)+(f^{\prime }(x)+b_{0}^{\prime }(x))^{2}/F\right] $, the
optimal $b_{0}(x)$ can minimize $\int_{0}^{a}G(b_{0},x)dx$, thereby
satisfying the Euler--Lagrange equation, i.e.,
\begin{equation}
\frac{\partial G(b_{0},x)}{\partial b_{0}(x)}-\frac{\partial }{\partial x}%
\frac{\partial G(b_{0},x)}{\partial b_{0}^{\prime }(x)}=0,  \label{7}
\end{equation}%
and the Neumann boundary condition%
\begin{equation}
\left. \frac{\partial G(b_{0},x)}{\partial b_{0}^{\prime }(x)}\right \vert
_{x=0}=\left. \frac{\partial G(b_{0},x)}{\partial b_{0}^{\prime }(x)}\right
\vert _{x=a}=0.  \label{8}
\end{equation}%
Based on Eqs. (\ref{7}) and (\ref{8}), one can finally obtain
\begin{equation}
p(x)b_{0}(x)=\frac{\partial }{\partial x}\left[ p(x)(f^{\prime
}(x)+b_{0}^{\prime }(x))/F\right] ,  \label{9}
\end{equation}%
with the boundary conditions $b_{0}^{\prime }(0)=-f^{\prime }(0)$ and $%
b_{0}^{\prime }(a)=-f^{\prime }(a).$ It is worth mentioning that Eq. (\ref{9}%
) can be solved either numerically or analytically to get the optimal bias $%
b_{0}(x)$ which can be substituted into Eq. (\ref{6}) for achieving the
valid lower bound eventually. Further, if the quantum Fisher information $F$%
\ is independent of $x$\ and the prior probability density $p(x)$\ is a
uniform distribution at\ $0\leq x\leq a$\ and $f(x)\equiv x$, one can thus
derive an analytical solution of Eq. (\ref{9}) with respect to the optimal
bias $b_{0}(x)$, i.e.,%
\begin{equation}
b_{0}(x)_{opt}=\frac{\cosh [\sqrt{F}(a-x)]-\cosh (\sqrt{F}x)}{\sqrt{F}\sinh
(a\sqrt{F})}.  \label{10}
\end{equation}%
By substituting Eqs. (\ref{10}) into (\ref{6}), as a consequence, one can
achieve the QOBB%
\begin{equation}
\Sigma \geq \Sigma _{O_{0}}=\frac{1}{F}-\frac{2}{aF^{3/2}}\tanh (a\sqrt{F}%
/2).  \label{11}
\end{equation}%
Due to the second term of Eq. (\ref{11}), we can easily see that the QOBB is
lower in numerical value than conventional QCRB. Moreover, when $a\sqrt{F}/2$
$\geq 2,$\ the Eq. (\ref{11}) can reduce to a more elegant result, i.e., $%
\Sigma _{O_{0}}\approx 1/F-2/aF^{3/2},$ since $\tanh (a\sqrt{F}/2)\approx 1$%
\ when $a\sqrt{F}/2$\ $\geq 2.$\ Then, we also consider some special
situations. For instance, when $F\rightarrow 0,$\ $\Sigma _{O_{0}}=a^{2}/12$%
\ is a non-negative value; while $a\rightarrow \infty ,$ both\ $\Sigma
_{O_{0}}=1/F$\ and QCRB are numerically identical. Additionally, when $%
a\rightarrow 0$\ or $F\rightarrow \infty ,$\ $\Sigma _{O_{0}}=0$\ represents
the trivial bound which can not provide any information.

\section{Two novel versions of QOBB under the unitary evolution process}

In this section, we mainly focus on considering two new quantization
processes to derive two novel versions of QOBB (respectively denoted as
QOBB1 and QOBB2) under an arbitrary unitary evolution process, such as a
phase encoding process and a SU(2) interferometer one, which can be seen in
the following section. For this purpose, the QOBB1 and the QOBB2 are
respectively presented from the viewpoint of the Cauchy-Schwarz inequality
and the variational method.

Let $x$ be the unknown parameter to be estimated and $\check{x}(y)$ be an
estimator of the unknown parameter $x$ from the observation results $y$. The
expectation value of the estimator $\check{x}(y)$ with respect to the
condition observation probability density $p(y|x)$ can be given by
\begin{eqnarray}
E_{y|x}[\check{x}(y)] &=&\int \check{x}(y)p(y|x)dy  \notag \\
&=&x+b_{1}(x),  \label{12}
\end{eqnarray}%
where $b_{1}(x)$ is the bias of the estimator $\check{x}(y)$. Note that Eq. (%
\ref{12}) can be also rewritten as%
\begin{equation}
\int [\check{x}(y)-E(x)]\text{Tr}(\hat{\rho}_{x}\hat{\Pi}_{y})dy=0,
\label{13}
\end{equation}%
where $E(x)\equiv E_{y|x}[\check{x}(y)]$ only depends on the unknown
parameter $x$. The corresponding mean-square estimation error can be
expressed as%
\begin{eqnarray}
\Sigma &=&\int (\check{x}(y)-x)^{2}p(x,y)dxdy  \notag \\
&=&\int p(x)(\delta \check{x}^{2}+b_{1}^{2}(x))dx,  \label{14}
\end{eqnarray}%
where $\delta \check{x}^{2}=\int (\check{x}(y)-E(x))^{2}$Tr$(\hat{\rho}_{x}%
\hat{\Pi}_{y})dy$ is the variance of the estimator $\check{x}(y)$. Now, let
us assume that the unknown parameter $x$ is mapped onto the quantum state $%
\hat{\rho}_{x}$ by the following unitary evolution process
\begin{equation}
\hat{\rho}_{x}=e^{i\hat{H}x}\hat{\rho}e^{-i\hat{H}x},  \label{15}
\end{equation}%
where $\hat{\rho}$ is the initial state and $\hat{H}$ is an Hermitian
operator that is independent of the unknown parameter $x$. In this
situation, differentiating Eq. (\ref{13}) with respect to $x$ and using Eq. (%
\ref{4}), one can obtain%
\begin{equation}
\int [\check{x}(y)-E(x)]\text{Tr}\left( \frac{\partial \hat{\rho}_{x}}{%
\partial x}\hat{\Pi}_{y}\right) dy=E^{\prime }(x).  \label{16}
\end{equation}%
Multiplying both sides of Eq. (\ref{16}) by the prior probability density $%
p(x)$ and invoking%
\begin{equation}
\frac{\partial \hat{\rho}_{x}}{\partial x}=-i(\hat{\rho}_{x}\hat{H}-\hat{H}%
\hat{\rho}_{x}),  \label{17}
\end{equation}%
which can be derived based on the Eq. (\ref{15}), one can find%
\begin{eqnarray}
&&\text{Im}\int p(x)(\check{x}(y)-E(x))\text{Tr}(\hat{\rho}_{x}\hat{H}\hat{%
\Pi}_{y})dy  \notag \\
&=&\frac{1}{2}E^{\prime }(x)p(x),  \label{18}
\end{eqnarray}%
with Im denoting the imaginary part. Further, we can multiply both sides of
Eq. (\ref{18}) by the real function $g(x)$ and then integrate with respect
to $x$, so that
\begin{equation}
\text{Im}\int \text{Tr}(\hat{A}^{\dagger }\hat{B})dxdy=\frac{1}{2}\int
p(x)E^{\prime }(x)g(x)dx,  \label{19}
\end{equation}%
where $\hat{A}^{\dagger }=\sqrt{p(x)}(\check{x}(y)-E(x))\sqrt{\hat{\Pi}_{y}}%
\sqrt{\hat{\rho}_{x}}$ and $\hat{B}=\sqrt{p(x)}g(x)\sqrt{\hat{\rho}_{x}}\hat{%
H}\sqrt{\hat{\Pi}_{y}}$. According to the Cauchy-Schwarz inequality, we see
that%
\begin{eqnarray}
&&\left \vert \text{Im}\int \text{Tr}(\hat{A}^{\dagger }\hat{B})dxdy\right
\vert ^{2}  \notag \\
&\leq &\left \vert \int \text{Tr}(\hat{A}^{\dagger }\hat{B})dxdy\right \vert
^{2}  \notag \\
&\leq &\left( \int \text{Tr}(\hat{A}^{\dagger }\hat{A})dxdy\right) \left(
\int \text{Tr}(\hat{B}^{\dagger }\hat{B})dxdy\right) ,  \notag \\
&=&\left( \int p(x)\delta \check{x}^{2}dx\right) \left( \int
p(x)g^{2}(x)\left \langle \hat{H}^{2}\right \rangle _{\rho }dx\right) ,
\label{20}
\end{eqnarray}%
where $\left \langle \hat{H}^{2}\right \rangle _{\rho }=$Tr$\left( \rho
H^{2}\right) $ is the average value\textbf{\ }for the initial state $\hat{%
\rho}$. Based on Eqs. (\ref{19}) and (\ref{20}), we have that%
\begin{equation}
\int p(x)\delta \check{x}^{2}dx\geq \frac{\left \vert \int p(x)E^{\prime
}(x)g(x)dx\right \vert ^{2}}{4\int p(x)g^{2}(x)\left \langle \hat{H}%
^{2}\right \rangle _{\rho }dx},  \label{21}
\end{equation}%
which is always true for the arbitrary real function $g(x)$ that satisfies
the inequality $\int p(x)g^{2}(x)\left \langle \hat{H}^{2}\right \rangle
_{\rho }dx>0$. In particular, when assuming that $g(x)=E^{\prime
}(x)/\left
\langle \hat{H}^{2}\right \rangle _{\rho }$, Eq. (\ref{21}) can
be reduced to
\begin{eqnarray}
\int p(x)\delta \check{x}^{2}dx &\geq &\int p(x)\frac{E^{\prime 2}(x)}{%
4\left \langle \hat{H}^{2}\right \rangle _{\rho }}dx,  \notag \\
&=&\int p(x)\frac{(1+b_{1}^{\prime }(x))^{2}}{4\left \langle \hat{H}%
^{2}\right \rangle _{\rho }}dx.  \label{22}
\end{eqnarray}%
Hence, based on Eq. (\ref{14}), a novel lower bound for the mean-square
estimation error under a unitary evolution process can be derived as%
\begin{equation}
\Sigma \geq \int p(x)\left[ \frac{(1+b_{1}^{\prime }(x))^{2}}{4\left \langle
\hat{H}^{2}\right \rangle _{\rho }}+b_{1}^{2}(x)\right] dx.  \label{23}
\end{equation}%
In order to minimize the novel bound given in Eq. (\ref{23}), similarly,
adopting the variational principle to derive the optimal bias $%
b_{1}(x)_{opt_{\text{I}}}$ is necessary. Likewise, we also assume that the
prior probability density $p(x)$\ is limited at\textbf{\ }$0\leq x\leq a$
and $\Gamma (b_{1},x)=$\ $p(x)[b_{1}^{2}(x)+(1+b_{1}^{\prime
}(x))^{2}/4\left \langle \hat{H}^{2}\right \rangle _{\rho }]$, the
corresponding optimal bias $b_{1}(x)_{opt_{\text{I}}}$ should satisfy the
Euler--Lagrange equation and the Neumann boundary condition like Eqs. (\ref%
{7}) and (\ref{8}), resulting in
\begin{equation}
p(x)b_{1}(x)=\frac{\partial }{\partial x}\left[ p(x)(1+b_{1}^{\prime
}(x))/4\left \langle \hat{H}^{2}\right \rangle _{\rho }\right] ,  \label{24}
\end{equation}%
with the boundary conditions $b_{1}^{\prime }(0)=b_{1}^{\prime }(a)=-1$. To
solve the analytic solution of Eq. (\ref{24}), we further suppose that the
prior probability density $p(x)$\ is a uniform distribution at the rang of $%
0\leq x\leq a$, so that%
\begin{equation}
b_{1}(x)_{opt_{\text{I}}}=\frac{\cosh [\sqrt{\Xi }(a-x)]-\cosh (\sqrt{\Xi }x)%
}{\sqrt{\Xi }\sinh (a\sqrt{\Xi })},  \label{25}
\end{equation}%
where $\Xi =4\left \langle \hat{H}^{2}\right \rangle _{\rho }.$

Upon substituting Eq. (\ref{25}) into Eq. (\ref{23}), the QOBB1 under the
unitary evolution process is given\ by
\begin{equation}
\Sigma \geq \Sigma _{O_{\text{I}}}=\frac{1}{\Xi }-\frac{2}{a\Xi ^{3/2}}\tanh
(a\sqrt{\Xi }/2).  \label{26}
\end{equation}%
It is worth noting that the value of QOBB1 is lower than $\Sigma _{O_{0}}$\
since $F\leq 4\left[ \left \langle \hat{H}^{2}\right \rangle _{\rho }-\left
\langle \hat{H}\right \rangle _{\rho }^{2}\right] \leq \Xi $\  \cite%
{50,51,52,53,54}$.$

After finishing the derivation of QOBB1, next we shall give another version
of QOBB, the QOBB2, following the results used by Ref. \cite{55}. To this
end, the non-Hermitian operator $\hat{Q}$ is introduced and then satisfies
\cite{55}
\begin{eqnarray}
\frac{\partial \hat{\rho}_{x}}{\partial x} &=&\frac{1}{2}(\hat{Q}\hat{\rho}%
_{x}+\hat{\rho}_{x}\hat{Q}^{\dagger }),  \notag \\
\text{Re}\left \langle \hat{Q}\right \rangle _{\hat{\rho}_{x}} &=&0,
\label{27}
\end{eqnarray}%
where Re represents the real part and $\left \langle \hat{Q}\right \rangle _{%
\hat{\rho}_{x}}=$Tr$(\hat{\rho}_{x}\hat{Q})$ is the average value for
quantum state $\hat{\rho}_{x}$. For a unitary evolution process described by
Eq. (\ref{15}), an appropriate choice for $\hat{Q}$ is the anti-Hermitian $%
\hat{Q}=2i\Delta \hat{H},$ where $\Delta \hat{H}=\hat{H}-\left \langle \hat{H%
}\right \rangle _{\hat{\rho}_{x}}$. Let $\hat{\rho}(x)=\hat{\rho}_{x}p(x)$
be the hybrid density operator in the hybrid quantum-classical space, which
also satisfies \cite{55}
\begin{equation}
\frac{\partial \hat{\rho}(x)}{\partial x}=\frac{1}{2}(\hat{L}(x)\hat{\rho}%
(x)+\hat{\rho}(x)\hat{L}^{\dagger }(x)),  \label{28}
\end{equation}%
with $\hat{L}(x)=\hat{Q}+\partial \ln p(x)/\partial x$. In a similar way to
derive Eq. (\ref{16}) and combine with Eq. (\ref{28}), one can also obtain%
\begin{equation}
\text{Re}\int (\check{x}(y)-E(x))\text{Tr}(\hat{\Pi}_{y}\hat{L}(x)\hat{\rho}%
(x))dy=E^{\prime }(x)p(x).  \label{29}
\end{equation}%
Thus, multiplying both sides of Eq. (\ref{29}) by the real function $h(x)$
and then integrating with respect to $x,$ we have that%
\begin{equation}
\text{Re}\int \text{Tr}(\hat{C}^{\dagger }\hat{D})dxdy=\int E^{\prime
}(x)p(x)h(x)dx,  \label{30}
\end{equation}%
where $\hat{C}^{\dagger }=h(x)\sqrt{\hat{\Pi}_{y}}\hat{L}(x)\sqrt{\hat{\rho}%
(x)}$ and $\hat{D}=(\check{x}(y)-E(x))\sqrt{\hat{\rho}(x)}\sqrt{\hat{\Pi}_{y}%
}$. Utilizing the Cauchy-Schwarz inequality again, one can finally find%
\begin{eqnarray}
&&\left \vert \text{Re}\int \text{Tr}(\hat{C}^{\dagger }\hat{D})dxdy\right
\vert ^{2}  \notag \\
&\leq &\left \vert \int \text{Tr}(\hat{C}^{\dagger }\hat{D})dxdy\right \vert
^{2}  \notag \\
&\leq &\left( \int \text{Tr}(\hat{C}^{\dagger }\hat{C})dxdy\right) \left(
\int \text{Tr}(\hat{D}^{\dagger }\hat{D})dxdy\right)  \notag \\
&=&\left( \int p(x)h^{2}(x)\Lambda dx\right) \left( \int p(x)\delta \check{x}%
^{2}dx\right) ,  \label{31}
\end{eqnarray}%
where $\Lambda =4\left[ \left \langle \hat{H}^{2}\right \rangle _{\rho
}-\left \langle \hat{H}\right \rangle _{\rho }^{2}\right] +\left( \partial
\lbrack \ln p(x)]/\partial x\right) ^{2}$. According to Eqs. (\ref{30}) and (%
\ref{31}), as a consequence, one can obtain%
\begin{equation}
\int p(x)\delta \check{x}^{2}dx\geq \frac{\left \vert \int E^{\prime
}(x)p(x)h(x)dx\right \vert ^{2}}{\int p(x)h^{2}(x)\Lambda dx},  \label{32}
\end{equation}%
which is valid for the arbitrary real function $h(x)$ with satisfying $\int
p(x)h^{2}(x)\Lambda dx>0$. In particular, when $h(x)=E^{\prime }(x)/\Lambda $%
, we get

\begin{eqnarray}
\int p(x)\delta \check{x}^{2}dx &\geq &\int p(x)\frac{E^{\prime 2}(x)}{%
\Lambda }dx  \notag \\
&=&\int p(x)\frac{(1+b_{1}^{\prime }(x))^{2}}{\Lambda }dx,  \label{33}
\end{eqnarray}%
so that according to Eq. (\ref{14}), we can finally derive another lower
bound for the mean-square estimation error%
\begin{equation}
\Sigma \geq \int p(x)\left[ \frac{(1+b_{1}^{\prime }(x))^{2}}{\Lambda }%
+b_{1}^{2}(x)\right] dx.  \label{34}
\end{equation}

Likewise, we also need to find out the optimal bias $b_{1}(x)_{opt_{II}}$
for minimizing the another lower bound shown in Eq. (\ref{32}). Analogously,
we set that the prior probability density $p(x)$\ is at\textbf{\ }$0\leq
x\leq a$ and $\Delta (b_{1},x)=p(x)[b_{1}^{2}(x)+(1+b_{1}^{\prime
}(x))^{2}/\Lambda ]$, the corresponding optimal bias $b_{1}(x)_{opt_{\text{II%
}}}$ which can minimize $\int_{0}^{a}$ $\Delta (b_{1},x)dx$, complying with
the Euler--Lagrange equation%
\begin{equation}
\frac{\partial \Delta (b_{1},x)}{\partial b_{1}(x)}-\frac{\partial }{%
\partial x}\frac{\partial \Delta (b_{1},x)}{\partial b_{1}^{\prime }(x)}=0,
\label{35}
\end{equation}%
with the Neumann boundary condition $\left. \partial \Delta
(b_{1},x)/\partial b_{1}^{\prime }(x)\right \vert _{x=0}=\left. \partial
\Delta (b_{1},x)/\partial b_{1}^{\prime }(x)\right \vert _{x=a}=0$, so that

\begin{equation}
p(x)b_{1}(x)=\frac{\partial }{\partial x}\left[ p(x)(1+b_{1}^{\prime
}(x))/\Lambda \right] ,  \label{36}
\end{equation}%
with the boundary conditions $b_{1}^{\prime }(0)=b_{1}^{\prime }(a)=-1$.
From Eq. (\ref{36}), if we further assume that the prior probability density%
\textbf{\ }$p(x)$\textbf{\ }is a uniform distribution at\textbf{\ }$0\leq
x\leq a$ and $\Lambda =4\left[ \left \langle \hat{H}^{2}\right \rangle
_{\rho }-\left \langle \hat{H}\right \rangle _{\rho }^{2}\right] $, then one
can obtain%
\begin{equation}
b_{1}(x)_{opt_{\text{II}}}=\frac{\cosh [\sqrt{\Lambda }(a-x)]-\cosh (\sqrt{%
\Lambda }x)}{\sqrt{\Lambda }\sinh (\sqrt{\Lambda }a)}.  \label{37}
\end{equation}%
Substituting Eq. (\ref{37}) into Eq. (\ref{34}), the QOBB2 under the unitary
evolution process can be given by%
\begin{equation}
\Sigma \geq \Sigma _{O_{\text{II}}}=\frac{1}{\Lambda }-\frac{2}{a\Lambda
^{3/2}}\tanh (a\sqrt{\Lambda }/2).  \label{38}
\end{equation}%
It is clearly seen that when the initial state $\hat{\rho}$\ is a mixed
state, the $\Sigma _{O_{\text{II}}}$\ is lower in numerical value than $%
\Sigma _{O_{0}}$\ since $F\leq $\ $\Lambda $\  \cite{50,51}, but if the
initial state $\hat{\rho}$\ is a pure state, then $\Sigma _{O_{\text{II}%
}}=\Sigma _{O_{0}}.$ Furthermore, since\textbf{\ }$\Lambda \leq \Xi ,$%
\textbf{\ }$\Sigma _{O_{\text{I}}}$\textbf{\ }$\leq $\textbf{\ }$\Sigma _{O_{%
\text{II}}}$ is also true.

As a consequence, here we give the relationship between these three kinds of
QOBB, i.e.,
\begin{equation}
\Sigma _{O_{\text{I}}}\leq \Sigma _{O_{\text{II}}}\leq \Sigma _{O_{0}}.
\label{39}
\end{equation}%
It is noteworthy from Eq. (\ref{39}) that the first inequality can be
saturated by using some special unitary evolution processes and initial
states (See more details in Sec. IV), while the second one can be saturated
by using initial pure states.

Before the end of this section, we need to consider the saturability problem
for the QOBB1 and the QOBB2. From the derivation process of Eq. (\ref{20}),
it can be seen that the QOBB1 can be fully saturated when Re[Tr$(\hat{\rho}%
_{x}\hat{H}\hat{\Pi}_{y})$]$=0$\ and the Cauchy-Schwarz inequality is
saturated, i.e, $\hat{A}\propto \hat{B}$ \cite{17,18,41}$,$\ which means
that $\left. \sqrt{\hat{\rho}_{x}}\sqrt{\hat{\Pi}_{y}}\right/ $Tr$(\hat{\rho}%
_{x}\hat{\Pi}_{y})$\ $=\left. \sqrt{\hat{\rho}_{x}}\hat{H}\sqrt{\hat{\Pi}_{y}%
}\right/ $Tr$(\hat{\rho}_{x}\hat{H}\hat{\Pi}_{y}).$\ Likewise, we can see
from Eq. (\ref{31}) that the QOBB2 can be completely saturated when Im[Tr$(%
\hat{\Pi}_{y}\hat{L}(x)\hat{\rho}(x))$]$=0$\ and $\left. \sqrt{\hat{\rho}(x)}%
\sqrt{\hat{\Pi}_{y}}\right/ $Tr$(\hat{\rho}(x)\hat{\Pi}_{y})=\left. \sqrt{%
\hat{\rho}(x)}\hat{L}^{\dagger }(x)\sqrt{\hat{\Pi}_{y}}\right/ $Tr$(\hat{\rho%
}(x)\hat{L}^{\dagger }(x)\hat{\Pi}_{y})$ hold simultaneously. However, since
$\hat{H}$ is an Hermitian operator and $\hat{L}(x)$ is an anti-Hermitian
operator, neither Re[Tr$(\hat{\rho}_{x}\hat{H}\hat{\Pi}_{y})$] nor Im[Tr$(%
\hat{\Pi}_{y}\hat{L}(x)\hat{\rho}(x))$] is equal to zero, which implies that
the QOBB1 and the QOBB2 cannot always be saturated. On the other hand, the
original QOBB has also similar problem, i.e., there is no fully general
guaranteed that we can saturate this bound in the non-asymptotic regime with
a limited number of measurements \cite{30}. Despite this, these three kinds
of QOBB can still be saturated in some special cases, such as the $N$\ spins
in the NOON state \cite{47} and the asymptotic regime with multiple
measurements (see Appendix B). Moreover, it is noteworthy that the
computational complexity of the QOBB1 and the QOBB2 is less than original
QOBB and QCRB in the case of initial mixed state since both of these two
error bounds involve the calculation of quantum Fisher information.

\section{\protect \bigskip Examples}

To intuitively see the estimation performance of these three kinds of QOBB,
in the following, we shall consider two specific cases of the unitary
evolution process, i.e., the phase encoding \cite{33,39} and SU(2)
interferometer processes \cite{56,57,58,59}.

Now, let us start with the phase encoding process of the quantum state $\hat{%
\rho}$ to an unknown phase $\phi ,$ i.e.,%
\begin{equation}
\hat{\rho}_{\phi }=e^{i\hat{n}\phi }\hat{\rho}e^{-i\hat{n}\phi },  \label{40}
\end{equation}%
where $\hat{n}=\hat{a}^{\dagger }\hat{a}$ is the photon number operator.
Notice that this phase encoding is indeed the conventional phase estimation
problem in quantum metrology. In this context, when given the coherent state
(CS) $\hat{\rho}_{\alpha }$ with $\alpha =\left \vert \alpha \right \vert
e^{i\theta _{\alpha }}$ ($\theta _{\alpha }=0$ for simplicity) and the
single-mode squeezed vacuum state (SMSVS) $\hat{\rho}_{r}$, we examine the
corresponding behaviors of these three kinds of QOBB. According to Eqs. (\ref%
{11}), (\ref{26}), and (\ref{38}), these three kinds of QOBB for the given
quantum states can be respectively derived as

\begin{eqnarray}
\Sigma _{O_{0}(\alpha )} &=&\frac{1}{4N_{\alpha }}-\frac{\tanh (a\sqrt{%
N_{\alpha }})}{4aN_{\alpha }^{3/2}},  \notag \\
\Sigma _{O_{\text{I}}(\alpha )} &=&\frac{1}{4N_{\alpha }(N_{\alpha }+1)}-%
\frac{\tanh [a\sqrt{N_{\alpha }(N_{\alpha }+1)}]}{4a\left[ N_{\alpha
}(N_{\alpha }+1)\right] ^{3/2}},  \notag \\
\Sigma _{O_{\text{II}}(\alpha )} &=&\Sigma _{O_{0}(\alpha )},  \notag \\
\Sigma _{O_{0}(r)} &=&\frac{1}{8N_{r}(N_{r}+1)}-\frac{\tanh [a\sqrt{%
2N_{r}(N_{r}+1)}]}{4a\left[ 2N_{r}(N_{r}+1)\right] ^{3/2}},  \notag \\
\Sigma _{O_{\text{I}}(r)} &=&\frac{1}{4N_{r}(2+3N_{r})}-\frac{\tanh [a\sqrt{%
N_{r}(2+3N_{r})}]}{4a[N_{r}(2+3N_{r})]^{3/2}},  \notag \\
\Sigma _{O_{\text{II}}(r)} &=&\Sigma _{O_{0}(r)},  \label{41}
\end{eqnarray}%
where $N_{\alpha }=\left \vert \alpha \right \vert ^{2}$ is the mean photon
number of $\hat{\rho}_{\alpha }$ and $N_{r}=\sinh ^{2}r$ is the mean photon
number of $\hat{\rho}_{r}$.

Then, let us examine the estimation performance of the MMSE. For this
purpose, by choosing the measurements in the bases of $\hat{\Pi}%
_{0}=\left
\vert \psi _{0}\right \rangle \left \langle \psi
_{0}\right
\vert $\ with $\left \vert \psi _{0}\right \rangle =(\left \vert
0\right
\rangle +\left
\vert 2\right \rangle )\left/ \sqrt{2}\right. $\
and $\hat{\Pi}_{1}=\hat{I}-\left \vert \psi _{0}\right \rangle \left \langle
\psi _{0}\right \vert $\ with $\hat{I}$\ being the identity operator.
According to Eq. (\ref{4}), we can respectively obtain the condition
observation probability densities for CS and SMSVS under the single-shot
measurement case, i.e.,
\begin{eqnarray}
p(k|\phi )_{(\alpha )} &=&C_{1}^{k}p^{k}(0|\phi )_{(\alpha )}p^{1-k}(1|\phi
)_{(\alpha )},  \notag \\
p(k|\phi )_{(r)} &=&C_{1}^{k}p^{k}(0|\phi )_{(r)}p^{1-k}(1|\phi )_{(r)},
\label{42}
\end{eqnarray}%
where $k\in \{0,1\}$\ and\textbf{\ }%
\begin{eqnarray}
p(0|\phi )_{(\alpha )} &=&e^{-\left \vert \alpha \right \vert ^{2}}(1+\left
\vert \alpha \right \vert ^{4}/2+\sqrt{2}\left \vert \alpha \right \vert
^{2}\cos 2\phi )/2,  \notag \\
p(1|\phi )_{(\alpha )} &=&1-p(0|\phi )_{(\alpha )},  \notag \\
p(0|\phi )_{(r)} &=&[1+(\tanh ^{2}r)/2  \notag \\
&&-\sqrt{2}(\tanh r)\cos 2\phi ]\left/ (2\cosh r)\right. ,  \notag \\
p(1|\phi )_{(r)} &=&1-p(0|\phi )_{(r)}.  \label{43}
\end{eqnarray}%
Substituting Eq. (\ref{42}) into Eq. (\ref{2}), one can finally derive the
MMSE for the given quantum states\textbf{\ }%
\begin{eqnarray}
MMSE_{(\alpha )} &=&\frac{a^{2}}{3}-\sum_{k=0}^{1}p(k)_{(\alpha )}\check{\phi%
}(k)_{opt(\alpha )}^{2},  \notag \\
MMSE_{(r)} &=&\frac{a^{2}}{3}-\sum_{k=0}^{1}p(k)_{(r)}\check{\phi}%
(k)_{opt(r)}^{2}.  \label{44}
\end{eqnarray}%
The specific forms of MMSE are shown in Appendix A and not shown here, for
simplicity.
\begin{figure}[tbp]
\label{Fig1} \centering \includegraphics[width=0.9\columnwidth]{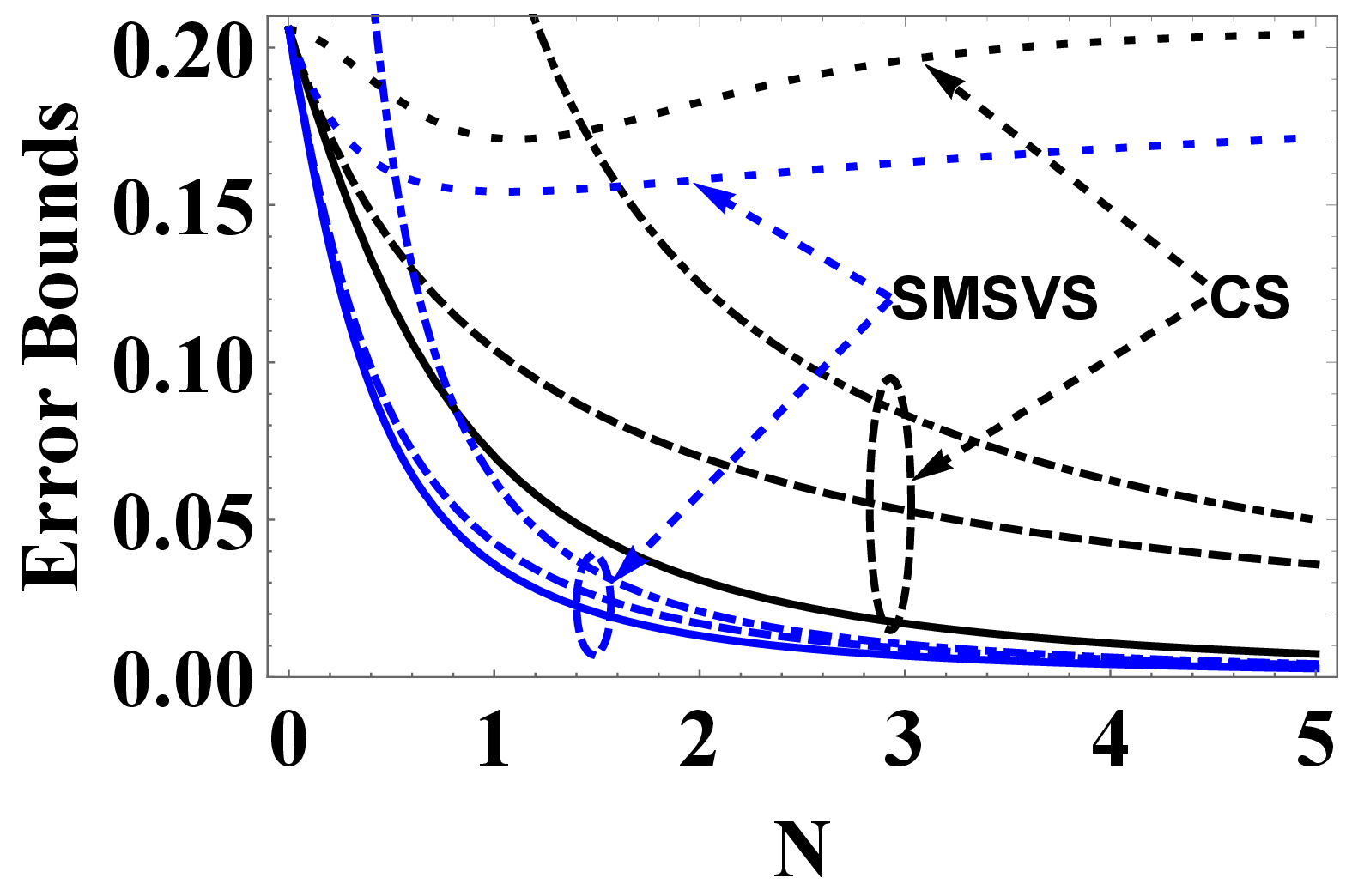}
\caption{{}(Color online) Error bounds as a function of the mean photon
number $N$ of the quantum state. The black and blue lines correspond to CS
and SMSVS, respectively.\textbf{\ }The dotted, dot-dashed, dashed and solid
lines correspond to the MMSE, QCRB, $\Sigma _{O_{0}}$(or $\Sigma _{O_{\text{%
II}}}$), and $\Sigma _{O_{\text{I}}}$\ respectively. }
\end{figure}
\begin{figure}[tbp]
\label{Fig2} \centering \includegraphics[width=0.9\columnwidth]{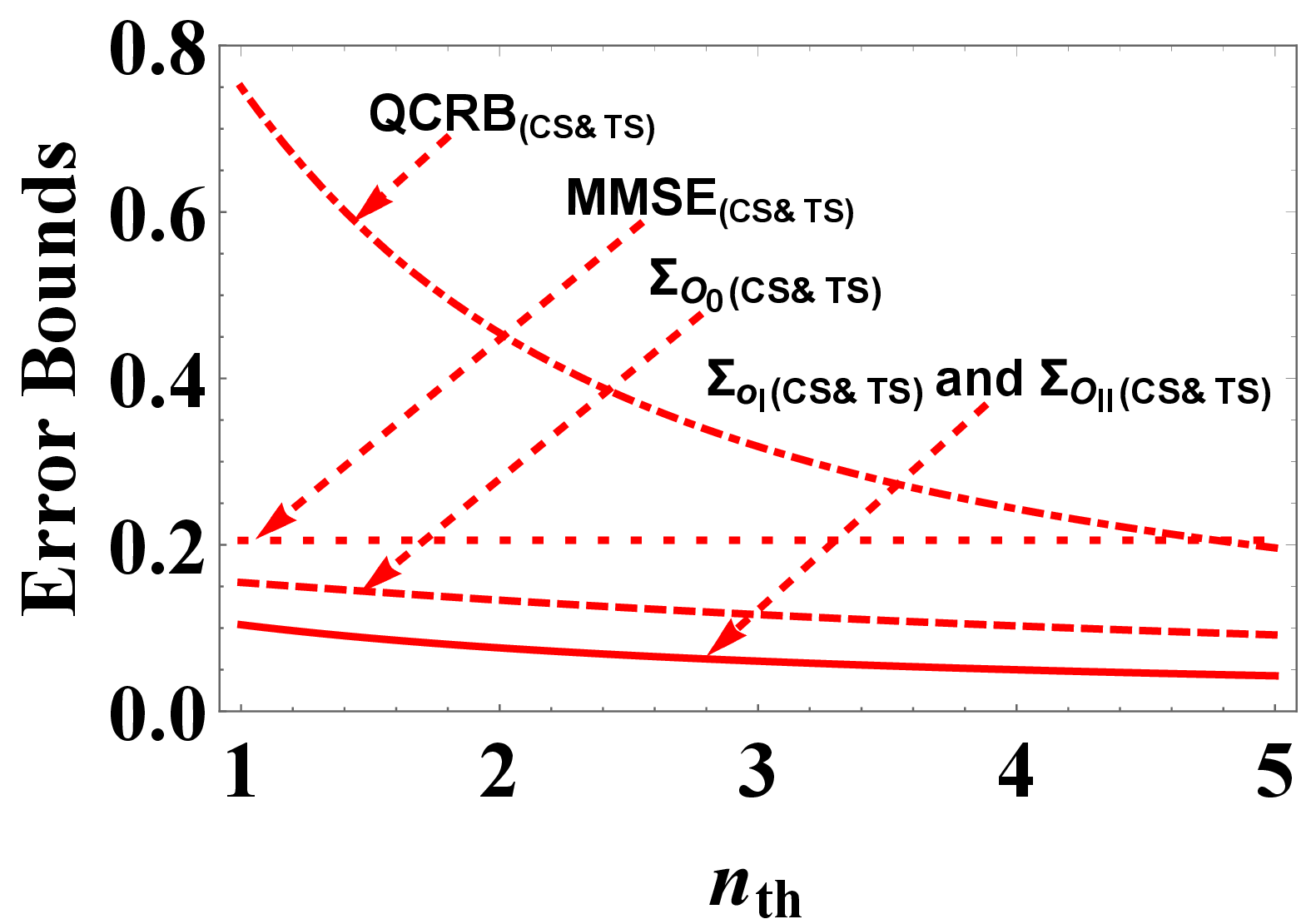}
\caption{{}(Color online) Error bounds as a function of the mean photon
number $N$ of the TS with $\left \vert \protect \beta \right \vert =1$. }
\end{figure}

In order to visually see the behaviors of these three kinds of QOBB, QCRB
and MMSE in phase estimation, at a fixed value of $a=\pi /2,$\ we plot these
error bounds as a function of the mean photon number $N$\ when given quantum
states, involving CS (black lines) and SMSVS (blue lines), as shown in Fig.
1. It is clear that the numerical value of SMSVS with respect to MMSE, QCRB,
$\Sigma _{O_{0}}$, $\Sigma _{O_{\text{II}}}$,\ and $\Sigma _{O_{\text{I}}}$\
is always lower than the cases of the CS. The reason may be that the SMSVS
is more nonclassical than the CS, so that the estimation performance of the
former shows better than that of the latter \cite{12}. In addition, since
both CS and SMSVS are pure states, this results in the same error bounds for
$\Sigma _{O_{0}}$\ and $\Sigma _{O_{\text{II}}}$, as expected. In
particular, for both CS and SMSVS, the corresponding $\Sigma _{O_{\text{I}}}$%
\ is the lowest in terms of numerical value, followed by $\Sigma
_{O_{0}}(\Sigma _{O_{\text{II}}})$, QCRB and MMSE. However, the tightness of
the $\Sigma _{O_{\text{I}}}$\ for both CS and SMSVS is the worst, followed
by $\Sigma _{O_{0}}(\Sigma _{O_{\text{II}}})$, QCRB since the gap between
the error bound and MMSE denotes the tightness, which also implies that the
larger the gap, the worse the tightness. As mentioned in the Sec. III, there
is no general guarantee about that we can saturate these three kinds of
QOBB. Nonetheless, we still provide a special example that can saturate the $%
\Sigma _{O_{0}}$\ and $\Sigma _{O_{\text{II}}},$ which refers to Appendix B
for more details.

Next, let us consider the second specific case, i.e., the SU(2)
interferometer process. In this situation, Eq. (\ref{15}) can be concretely
written as \cite{60,61}
\begin{equation}
\hat{\rho}_{out}=e^{-i\theta \hat{J}_{y}}\hat{\rho}_{in}e^{i\theta \hat{J}%
_{y}},  \label{45}
\end{equation}%
where $\hat{\rho}_{in}$ and $\hat{\rho}_{out}$ are respectively the input
and output states of SU(2) interferometer systems, $\theta $\textbf{\ }is
the unknown parameter to be estimated, and $\hat{J}_{y}=-\frac{i}{2}(\hat{a}%
^{\dagger }\hat{b}-\hat{a}\hat{b}^{\dagger })$ with $\hat{a}$ \ ($\hat{a}%
^{\dagger }$) and $\hat{b}$ ($\hat{b}^{\dagger }$) being the annihilation
(addition) operators for modes $a$ and $b$, respectively. For the sake of
calculation, here we take both the CS $\hat{\rho}_{\beta }$ with $\beta
=\left \vert \beta \right \vert e^{i\theta _{\beta }}$ ($\theta _{\beta }=0$
for simplicity) and the thermal state (TS) $\hat{\rho}_{th}$ as the inputs%
\textbf{\ }in modes $a$\ and $b$, respectively. Thus, using Eqs. (\ref{11}),
(\ref{26}), and (\ref{38}), one can respectively obtain these three kinds of
QOBB, i.e.,%
\begin{eqnarray}
\Sigma _{O_{0}(CS\&TS)} &=&\frac{1}{F_{CS\&TS}}-\frac{2\tanh (a\sqrt{%
F_{CS\&TS}}/2)}{aF_{CS\&TS}^{3/2}},  \notag \\
\Sigma _{O_{\text{I}}(CS\&TS)} &=&\frac{1}{4\left \langle \hat{J}%
_{y}{}^{2}\right \rangle }-\frac{\tanh \left( a\sqrt{\left \langle \hat{J}%
_{y}{}^{2}\right \rangle }\right) }{4a\left \langle \hat{J}%
_{y}{}^{2}\right \rangle ^{3/2}},  \notag \\
\Sigma _{O_{\text{II}}(CS\&TS)} &=&\Sigma _{O_{\text{I}}(CS\&TS)},
\label{46}
\end{eqnarray}%
where $F_{CS\&TS}=\bar{n}_{th}+\left \vert \beta \right \vert ^{2}/(2\bar{n}%
_{th}+1)$ is the quantum Fisher information that can be seen from Ref. \cite%
{61} with the mean photon number of TS $\bar{n}_{th}$, and $\left \langle
\hat{J}_{y}{}^{2}\right \rangle =$ $(\left \vert \beta \right \vert ^{2}+%
\bar{n}_{th}+2\left \vert \beta \right \vert ^{2}\bar{n}_{th})/4$ is the
average value for the input states of SU(2) interferometer.

To further show the advantage of these three kinds of QOBB, here we present
the MMSE for the SU(2) interferometer process. To this end, we use $\hat{%
\lambda}_{0}=\left \vert 10\right \rangle \left \langle 01\right \vert $\
and $\hat{\lambda}_{1}=\hat{I}-\left \vert 10\right \rangle \left \langle
01\right \vert $\ as the measurement basis. Thus, using Eq. (\ref{4}), the
condition observation probability density is given by%
\begin{equation}
p(k|\theta )_{(CS\&TS)}=C_{1}^{k}p^{k}(0|\phi )_{(CS\&TS)}p^{1-k}(1|\phi
)_{(CS\&TS)},  \label{47}
\end{equation}%
where%
\begin{eqnarray}
&&p(0|\theta )_{(CS\&TS)}  \notag \\
&=&\frac{\left \vert \beta \right \vert ^{2}e^{-\left \vert \beta \right
\vert ^{2}}}{\bar{n}_{th}+1}\cos ^{2}\frac{\theta }{2}+\frac{\bar{n}%
_{th}e^{-\left \vert \beta \right \vert ^{2}}}{(\bar{n}_{th}+1)^{2}}\sin ^{2}%
\frac{\theta }{2},  \notag \\
&&p(1|\theta )_{(CS\&TS)}=1-p(0|\theta )_{(CS\&TS)}.  \label{48}
\end{eqnarray}%
By substituting Eq. (\ref{47}) into Eq. (\ref{2}), one can obain the MMSE
for the CS and TS under the SU(2) interferometer process%
\begin{equation}
MMSE_{(CS\&TS)}=\frac{a^{2}}{3}-\sum_{k=0}^{1}p(k)_{(CS\&TS)}\check{\theta}%
(k)_{opt(CS\&TS)}^{2}.  \label{49}
\end{equation}%
One can refer to the Appendix C for the specific expression of Eq. (\ref{49}%
).

To deeply understand the behaviors of these three kinds of QOBB, QCRB and
MMSE in SU(2) interferometer process, at fixed value of $a=\pi /2$\ and $%
\left \vert \beta \right \vert =1,$\ we also plot these error bounds
changing with $\bar{n}_{th}$, as shown in Fig. 2. It is found that both $%
\Sigma _{O_{\text{I}}}$\ and $\Sigma _{O_{\text{II}}}$\ not only have the
same numerical value, but also are lower than $\Sigma _{O_{0}}$, MMSE and
QCRB. The one reason for these phenomena is that because of the special
SU(2) interferometer process and the input state TS, this results in $\left
\langle \hat{J}_{y}{}\right \rangle =0$\ so that $\Sigma _{O_{\text{II}%
}(CS\&TS)}=\Sigma _{O_{\text{I}}(CS\&TS)}.$\ The other one is that, since TS
is a mixed state, this leads to that the value of $\Sigma _{O_{\text{II}%
}(CS\&TS)}$\ is lower than $\Sigma _{O_{0}(CS\&TS)}$\ case. From the
perspective of these error bounds' tightness, the $\Sigma _{O_{0}}$\ is
superior to both $\Sigma _{O_{\text{I}}}$\ and $\Sigma _{O_{\text{II}}}$\
since the gap between the $\Sigma _{O_{0}}$\ and MMSE is relatively small.

\section{Conclusions}

In summary, we consider two new quantization processes of the classical
optimal biased bound in the framework of a unitary evolution to derive the
other two versions of the QOBB, i.e., $\Sigma _{O_{\text{I}}}$\ and $\Sigma
_{O_{\text{II}}}$. Compared with the original $\Sigma _{O_{0}},$\ the $%
\Sigma _{O_{\text{I}}}$\ and $\Sigma _{O_{\text{II}}}$\ can offer lower
precision limit when the initial state is a mixed state. For the initial
pure state scenario, the value of $\Sigma _{O_{\text{I}}}$\ is still lower
than $\Sigma _{O_{0}}$, while the $\Sigma _{O_{\text{II}}}$\ and $\Sigma
_{O_{0}}$\ are the same in terms of numerical values. Nevertheless, the $%
\Sigma _{O_{\text{I}}}$\ and $\Sigma _{O_{\text{II}}}$\ can provide
precision limits which are not higher than original $\Sigma _{O_{0}},$ but
the tightness of the $\Sigma _{O_{\text{I}}}$\ and $\Sigma _{O_{\text{II}}}$%
\ is worse than that of the $\Sigma _{O_{0}}.$ Further, we have demonstrated
our results with two specific examples of the unitary evolution process,
including the phase encoding process and the SU(2) interferometer process.

Finally, we should mention that, although these three kinds of QOBB can
provide lower precision limit than QCRB$,$ there is no general guarantee
about that these three kinds of QOBB are fully tight. Therefore, finding a
new quantization process to derive a completely tight QOBB remains an open
problem.

\begin{acknowledgments}
This work was supported by the National Nature Science Foundation of China
(Grant Nos. 91536115, 11534008, 62161029); Natural Science Foundation of
Shaanxi Province (Grant No. 2016JM1005); Shaanxi Fundamental Science
Research Project of Mathematics and Physics (Grant No. 22JSY005); Natural
Science Foundation of Jiangxi Provincial (Grant No. 20202BABL202002). Wei Ye
is supported by both Jiangxi Provincial Natural Science Foundation
(20232BAB211032) and Scientific Research Startup Foundation (Grant No.
EA202204230) at Nan chang Hangkong University.
\end{acknowledgments}

\textbf{Appendix A: The} \textbf{specific expression of }Eq. (\ref{44})

Using Eqs. (\ref{2}) and (\ref{42}), one can respectively obtain the MMSE
for CS and SMSVS%
\begin{align}
MMSE_{(\alpha )}& =\frac{a^{2}}{3}-\sum_{k=0}^{1}p(k)_{(\alpha )}\check{\phi}%
(k)_{opt(\alpha )}^{2},  \notag \\
MMSE_{(r)}& =\frac{a^{2}}{3}-\sum_{k=0}^{1}p(k)_{(r)}\check{\phi}%
(k)_{opt(r)}^{2},  \tag{A1}
\end{align}%
where we have set\textbf{\ }%
\begin{widetext}
\begin{align}
& \sum_{k=0}^{1}p(k)_{(\alpha )}\check{\phi}(k)_{opt(\alpha )}^{2}=\frac{%
\left( \int_{0}^{a}\phi p(0|\phi )_{(\alpha )}d\phi \right) ^{2}}{%
a\int_{0}^{a}p(0|\phi )_{(\alpha )}d\phi }+\frac{\left( \int_{0}^{a}\phi
p(1|\phi )_{(\alpha )}d\phi \right) ^{2}}{a\int_{0}^{a}p(1|\phi )_{(\alpha
)}d\phi },  \notag \\
& \sum_{k=0}^{1}p(k)_{(r)}\check{\phi}(k)_{opt(r)}^{2}=\frac{\left(
\int_{0}^{a}\phi p(0|\phi )_{(r)}d\phi \right) ^{2}}{a\int_{0}^{a}p(0|\phi
)_{(r)}d\phi }+\frac{\left( \int_{0}^{a}\phi p(1|\phi )_{(r)}d\phi \right)
^{2}}{a\int_{0}^{a}p(1|\phi )_{(r)}d\phi },  \tag{A2}
\end{align}%
with%
\begin{align}
& \int_{0}^{a}p(0|\phi )_{(\alpha )}d\phi =e^{-\left \vert \alpha \right
\vert ^{2}}[(1+\left \vert \alpha \right \vert ^{4}/2)a+\left \vert \alpha
\right \vert ^{2}(\sin 2a)/\sqrt{2}]/2,  \notag \\
& \int_{0}^{a}\phi p(0|\phi )_{(\alpha )}d\phi =e^{-\left \vert \alpha
\right \vert ^{2}}[(1+\left \vert \alpha \right \vert ^{4}/2)a^{2}+\sqrt{2}%
\left \vert \alpha \right \vert ^{2}(a\sin 2a+(\cos 2a)/2-1/2)]/4,  \notag \\
& \int_{0}^{a}p(1|\phi )_{(\alpha )}d\phi =a-\int_{0}^{a}p(0|\phi )_{(\alpha
)}d\phi ,  \notag \\
& \int_{0}^{a}\phi p(1|\phi )_{(\alpha )}d\phi =\frac{a^{2}}{2}%
-\int_{0}^{a}\phi p(0|\phi )_{(\alpha )}d\phi ,  \tag{A3}
\end{align}%
and%
\begin{align}
& \int_{0}^{a}p(0|\phi )_{(r)}d\phi =[(1+(\tanh ^{2}r)/2)a-(\tanh r)(\sin
2a)/\sqrt{2}]/(2\cosh r),  \notag \\
& \int_{0}^{a}\phi p(0|\phi )_{(r)}d\phi =[(1+(\tanh ^{2}r)/2)a^{2}-\sqrt{2}%
(\tanh r)(a\sin 2a+(\cos 2a)/2-1/2)]/(4\cosh r),  \notag \\
& \int_{0}^{a}p(1|\phi )_{(r)}d\phi =a-\int_{0}^{a}p(0|\phi )_{(r)}d\phi ,
\notag \\
& \int_{0}^{a}\phi p(1|\phi )_{(r)}d\phi =\frac{a^{2}}{2}-\int_{0}^{a}\phi
p(0|\phi )_{(r)}d\phi .  \tag{A4}
\end{align}%
\end{widetext}
\textbf{Appendix B: A special example that can saturate the original QOBB
and QOBB2.}

\bigskip In this appendix, we give a special example for the phase encoding
process that can saturate the original $\Sigma _{O_{0}}$\ and $\Sigma _{O_{%
\text{II}}}$. Specifically, we choose $\left \vert \varphi _{0}\right
\rangle =\left. (\left \vert 0\right \rangle +\left \vert 1\right \rangle
)\right/ \sqrt{2}$\ as the initial quantum state and $\hat{M}_{k}=\left
\vert \varphi _{k}\right \rangle \left \langle \varphi _{k}\right \vert $\ $%
(k=0,1)$\ with $\left \vert \varphi _{1}\right \rangle =\left. (\left \vert
0\right \rangle -\left \vert 1\right \rangle )\right/ \sqrt{2}$\ as the
measurement basis. Further, we consider repeating the same measurement $v$\
times and then get the condition observation probability density according
to Eq. (\ref{4}), which can be given by
\begin{equation}
p(m|\phi )_{(\left \vert \varphi _{0}\right \rangle
)}=C_{v}^{m}p_{0}^{m}p_{1}^{v-m},  \tag{B1}
\end{equation}%
\ where $C_{v}^{m}=v!/[m!(v-m)!]$\ is the binomial coefficient, $p_{0}=\cos
^{2}(\phi /2),$\ and $p_{1}=\sin ^{2}(\phi /2).$\ Substituting Eq. (B1) into
Eq. (\ref{2}), one can theoretically derive the corresponding MMSE for the
given initial quantum state $\left \vert \varphi _{0}\right \rangle $, not
show here for simplicity. Besides, as a comparison, we also derive three
kinds of QOBB for the given $\left \vert \varphi _{0}\right \rangle $\
according to Eqs. (\ref{11}), (\ref{26}), and (\ref{38}), which can be
respectively expressed as%
\begin{align}
\Sigma _{O_{0}(\left \vert \varphi _{0}\right \rangle )}& =\frac{1}{v}-\frac{%
2}{av^{3/2}}\tanh \left( a\sqrt{v}/2\right) ,  \notag \\
\Sigma _{O_{\text{I}}(\left \vert \varphi _{0}\right \rangle )}& =\frac{1}{2v%
}-\frac{2}{a(2v)^{3/2}}\tanh (a\sqrt{2v}/2),  \notag \\
\Sigma _{O_{\text{II}}(\left \vert \varphi _{0}\right \rangle )}& =\Sigma
_{O_{0}(\left \vert \varphi _{0}\right \rangle )}.  \tag{B2}
\end{align}%
\begin{figure}[tbp]
\label{Fig3} \centering \includegraphics[width=0.9\columnwidth]{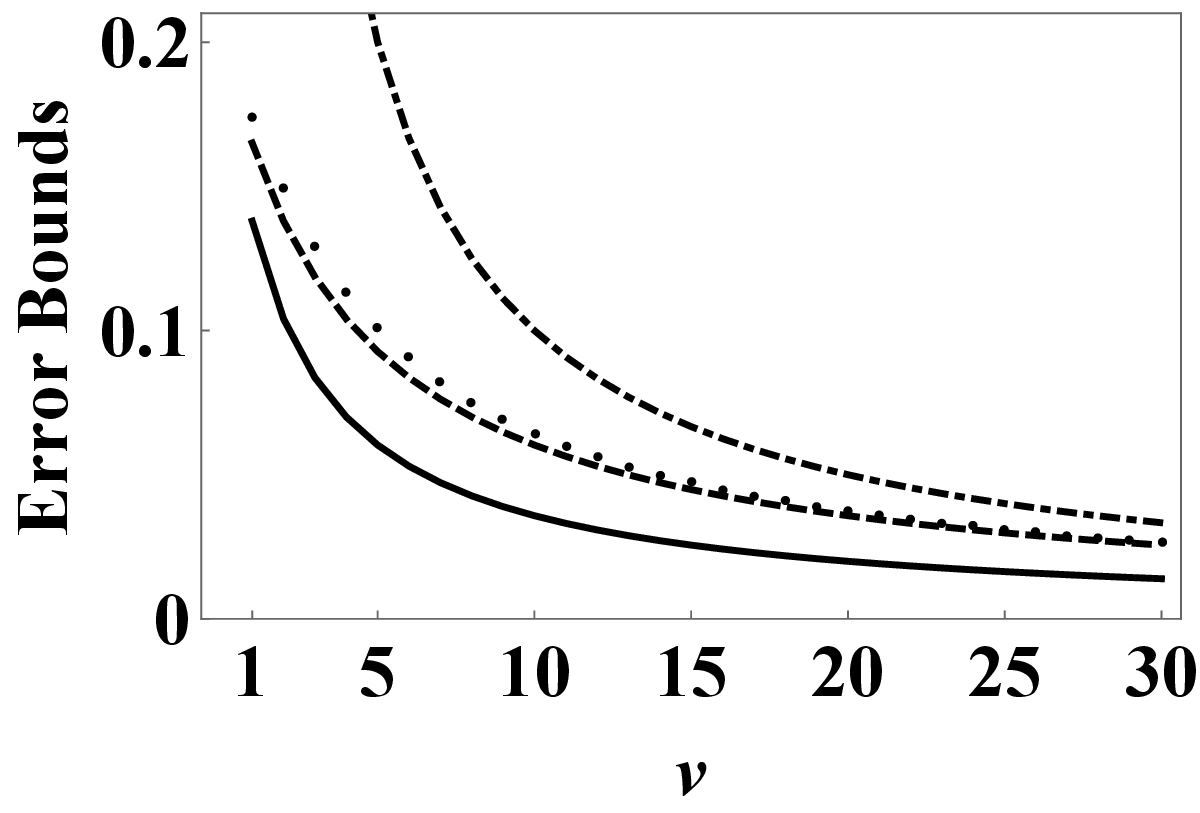}
\caption{{}(Color online) Error bounds as a function of the number of
measurement $v$. The dotted, dot-dashed, dashed and solid lines respectively
correspond to the MMSE, QCRB, $\Sigma _{O_{0}}$(or $\Sigma _{O_{\text{II}}}$%
), and $\Sigma _{O_{\text{I}}}$. }
\end{figure}
In order to visually show the estimation performance of the MMSE, QCRB and
these three kinds of QOBB, at a fixed value of $a=\pi /2,$\ we plot these
error bounds as a function of $v$, as shown in Fig. 3. The results show that
when the number of measurement $v$\ is relatively large, the MMSE can attain
the original $\Sigma _{O_{0}}$\ and $\Sigma _{O_{\text{II}}}.$

\textbf{Appendix C: The} \textbf{specific expression of }Eq. (\ref{49})

According to Eqs. (\ref{2}) and (\ref{47}), one can derive the MMSE for the
CS and TS under the SU(2) interferometer process%
\begin{equation}
MMSE_{(CS\&TS)}=\frac{a^{2}}{3}-\sum_{k=0}^{1}p(k)_{(CS\&TS)}\check{\theta}%
(k)_{opt(CS\&TS)}^{2},  \tag{C1}
\end{equation}%
where we have set%
\begin{widetext}
\begin{equation}
\sum_{k=0}^{1}p(k)_{(CS\&TS)}\check{\theta}(k)_{opt(CS\&TS)}^{2}=\frac{%
\left( \int_{0}^{a}\theta p(0|\theta )_{(CS\&TS)}d\theta \right) ^{2}}{%
a\int_{0}^{a}p(0|\theta )_{(CS\&TS)}d\theta }+\frac{\left(
\int_{0}^{a}\theta p(1|\theta )_{(CS\&TS)}d\theta \right) ^{2}}{%
a\int_{0}^{a}p(1|\theta )_{(CS\&TS)}d\theta },  \tag{C2}
\end{equation}%
with%
\begin{align}
& \int_{0}^{a}p(0|\theta )_{(CS\&TS)}d\theta =\frac{\left \vert \beta \right
\vert ^{2}e^{-\left \vert \beta \right \vert ^{2}}}{2\left( \bar{n}%
_{th}+1\right) }(a+\sin a)+\frac{\bar{n}_{th}e^{-\left \vert \beta \right
\vert ^{2}}}{2(\bar{n}_{th}+1)^{2}}(a-\sin a),  \notag \\
& \int_{0}^{a}\theta p(0|\theta )_{(CS\&TS)}d\theta =\frac{\left \vert \beta
\right \vert ^{2}e^{-\left \vert \beta \right \vert ^{2}}}{2\left( \bar{n}%
_{th}+1\right) }(a^{2}/2+a\sin a+\cos a-1)+\frac{\bar{n}_{th}e^{-\left \vert
\beta \right \vert ^{2}}}{2(\bar{n}_{th}+1)^{2}}\left[ a^{2}/2-\left( a\sin
a+\cos a-1\right) \right] ,  \notag \\
& \int_{0}^{a}p(1|\theta )_{(CS\&TS)}d\theta =a-\int_{0}^{a}p(0|\theta
)_{(CS\&TS)}d\theta ,  \notag \\
& \int_{0}^{a}\theta p(1|\theta )_{(CS\&TS)}d\theta =\frac{a^{2}}{2}%
-\int_{0}^{a}\theta p(0|\theta )_{(CS\&TS)}d\theta .  \tag{C3}
\end{align}
\end{widetext}

\end{document}